\documentclass{llncs}
\usepackage{amsfonts}
\usepackage{amscd}
\usepackage{amssymb}
\usepackage{pifont}
\usepackage{amsmath}
\usepackage{epsfig}
\begin{document}

\title{Authenticated Multiuser Quantum Direct Communication using Entanglement Swapping}%
\author{Changho Hong$^a$ \and Jiin Kim$^b$ \and Hwayean Lee$^a$ \and Hyungjin Yang$^{c,d}$}%
\institute{$^a$Center for Information Security Technologies(CIST),\\
           Korea University, Seoul, South Korea\\
$^b$Department of physics, Korea University, Seoul, South Korea\\
$^c$Department of physics, Korea University, Chochiwon, Choongnam, South Korea\\
$^d$Graduate School of Information Security(GSIS), \\Korea
University, 1, 5-ka, Anam-dong Sungbuk-ku, Seoul, South Korea\\
\email{hchc@korea.ac.kr, jiinny@korea.ac.kr,
hylee@cist.korea.ac.kr, yangh@korea.ac.kr}} \maketitle

\begin{abstract}
We present an Authenticated Multiuser Quantum Direct
Communication(MQDC) protocol using entanglement swapping. Quantum
direct communication is believed to be a safe way to send a secret
message without quantum key distribution.  The authentication
process in our protocol allows only proper users to participate in
communication.  In this communication stage after the
authentication, any two authorized users among n users can
communicate each other even though there is no quantum
communication channels between them.  In the protocol, we need
only n quantum communication channels between the authenticator
and n users.  It is similar to the present telephone system in
which there are n communication channels between telephone company
and users and any two designated users can communicate each other
using telephone line through the telephone company.  The
securities of our protocols are analysed to be the same as those
of other quantum key distribution protocols.
\end{abstract}

{\em Introduction}-One of the objects of quantum cryptography is to
allow two distant parties to share a random bit sequence without any
reveals to the eavesdropper.  The Quantum Key Distribution(QKD)
protocols are regarded as unconditionally secure cryptography
schemes.  The first Quantum key distribution was proposed by Bennett
and Brassard.  It is known as the BB84 protocol\cite{Bennett}, and
it uses four different non-orthogonal states of single photon. QKD
establishes a common random key between two remote parties of
communication. Afterwards these two parties can safely exchange a
secret message over the public channel by encoding and decoding them
with the distributed key. If the length of the keys is the same as
the length of the messages, the communication is unconditional
secure. It is because that one-time pad scheme with the enough
length of secret key is proved to be unconditionally secure.  QKD
has progressed quickly since the first QKD protocol was
designed\cite{Ekert,Bennett_2,golden,ard}. QKD based on quantum
mechanics is usually
non-deterministic\cite{Bennett,Ekert,Bennett_2,golden,ard}. But it
is sometimes deterministic\cite{Koashi,Bennett_4,Hwang}, in which
two remote parties get the same keys determinately.

A novel concept of quantum direct communication (QDC) has been
proposed and pursued recently.  Unlike QKD, QDC can directly send
secret messages without creating the key to encrypt them.  In 2002,
Beige et al. presented the first QDC scheme,\cite{Beige} in which
messages can be read after the transmission of classical
informations.  Bostrom and Felbinger put forward a ping-ping scheme
using entangled pair of qubits in 2002\cite{BF}.  This protocol can
be used for QKD as well as QDC.  It is secure for key distribution,
but is only quasisecure for QDC even if perfect quantum channel is
used.  Cai modified the ping-pong protocol by replacing the
entanglement states with single photons in mixed state\cite{cai}.
However it is unsafe in a noisy channel and disadvantaged to the
opaque attack.

QDC may have wide application due to its fastness and unconditional
security.  Our QDC protocol uses entangled states and the
entanglement swapping effect.  It is well known that quantum
entanglement swapping\cite{Zeil} can entangle two quantum systems
which did not interact with each other before. But these QDC
protocols have a common serious problem.  If we don't check whether
only proper users communicate each other, secret messages can be
exposed to the eavesdropper.  It is, thus, important to certify the
identifies of the legitimate users in communication line so that no
third party monitoring their identification can impersonate either
of them.  In our protocol, Alice (or Bob) can confirm the
identification of her (or his) counterpart through the trusted third
party, Trent, who acts a role of present telephone company.  When
one of them wants to communicate with the other, Trent guarantees
the identification of each person to his(her) counterpart.
Afterwards they directly communicate each other
using quantum communication channels linking them and Trent.

Our authenticated multiuser QDC scheme using entanglement swapping
consists of two parts; quantum authentication mode and quantum
communication mode.  After finishing authentication mode to identify
each other, the messages are transmitted secretly and directly in
communication mode.

{\em Entanglement swapping}-Let us first describe the quantum
entanglement swapping.  Let $|0 \rangle$ and $|1 \rangle$ be the
horizontal and vertical polarization states of a photon,
respectively.  The four Bell states, $|\Phi^{\pm} \rangle \equiv
\frac {1}{\sqrt {2}}( |00\rangle \pm |11\rangle)$ and $|\Psi^{\pm}
\rangle \equiv \frac {1}{\sqrt {2}}( |01\rangle \pm |10\rangle)$ are
maximally entangled states in two-photon Hilbert space. Let the
initial state is $|\Phi_{12}^{+} \rangle \otimes |\Phi_{34}
^{+}\rangle$. We can see that after the Bell measurements on the
pair of photon 1 and 3 and the pair of photon 2 and 4, there is an
explicit correspondence between the known initial state of the pair
of two qubits and its swapped measurement outcomes.  The state
$|\Phi_{12}^{+} \rangle \otimes |\Phi_{34} ^{+}\rangle$ can be
rearranged as the linear combinations of the terms, $|\Phi_{13}^{+}
\rangle \otimes |\Phi_{24} ^{+}\rangle$, $|\Phi_{13}^{-} \rangle
\otimes |\Phi_{24} ^{-}\rangle$, $|\Psi_{13}^{+} \rangle \otimes
|\Psi_{24} ^{+}\rangle$ and $|\Psi_{13}^{-} \rangle \otimes
|\Psi_{24} ^{-}\rangle$. When the outcome of Bell measurement on the
pair of photon 1 and 3 is $|\Phi_{13} ^{-}\rangle$, the Bell state
of the pair of photon 2 and 4 must be $|\Phi_{24} ^{-}\rangle$. The
outcome of entanglement swapping is summarized in Table \ref{out}.

In our protocol, every user sends Trent the secret identity sequence
of $N$-bits. We call the Alice's(Bob's) secret identity as $ID(A)
(ID(B))$. It must be kept safely between the user and Trent.  Let us
introduce the explicit algorithm for the protocol.

\begin{table}
\caption{\textbf{The outcomes of the swapped Bell measurement on the
initially different combinations of four Bell states} The
abbreviation ID++ represents the set of four possible outcomes of
Bell measurement,$(|\Phi_{14}^{+} \rangle
,|\Phi_{23}^{+}\rangle)$,$(|\Phi_{14}^{-} \rangle
,|\Phi_{23}^{-}\rangle)$,$(|\Psi_{14}^{+} \rangle
,|\Psi_{23}^{+}\rangle)$, and $(|\Psi_{14}^{-}
\rangle,|\Psi_{23}^{-}\rangle)$ with equal probability of 1/4.
Similarly, the following cases can be obtained. $ID+- \Rightarrow$
$\{(|\Psi_{14}^{+} \rangle,|\Psi_{23}^{-}\rangle)$,$(|\Phi_{14}^{-}
\rangle,|\Phi_{23}^{+}\rangle)$,$(|\Phi_{14}^{+}
\rangle,|\Phi_{23}^{-}\rangle)$,$(|\Psi_{14}^{-}
\rangle,|\Psi_{23}^{+}\rangle)\}$, $Rev++ \Rightarrow$
$\{(|\Phi_{14}^{+} \rangle,|\Psi_{23}^{+}\rangle )$,$(|\Phi_{14}^{-}
\rangle,|\Psi_{23}^{-}\rangle)$,$(|\Psi_{14}^{+}
\rangle,|\Phi_{23}^{+}\rangle)$,$(|\Psi_{14}^{-}
\rangle,|\Phi_{23}^{-}\rangle)\}$, and $Rev+- \Rightarrow$
$\{(|\Phi_{14}^{+} \rangle,|\Psi_{23}^{-}\rangle)$,$(|\Phi_{14}^{-}
\rangle,|\Psi_{23}^{+}\rangle)$,$(|\Psi_{14}^{+}
\rangle,|\Phi_{23}^{-}\rangle)$,$(|\Psi_{14}^{-}
\rangle,|\Phi_{23}^{+}\rangle)\}.$}
\begin{center}
\begin{tabular}{|c|p{2cm}p{2cm}p{2cm}c|} \hline
& $|\Phi_{34}^{+}\rangle$ & $|\Phi_{34}^{-}\rangle$ &
$|\Psi_{34}^{+}\rangle$ & $|\Psi_{34}^{-}\rangle$ \\ \hline \hline
$|\Phi_{12}^{+}\rangle$ & $ID++$ & $ID+-$ & $Rev++$ & $Rev+-$ \\
$|\Phi_{12}^{-}\rangle$ & $ID+-$ & $ID++$ & $Rev+-$ & $Rev++$ \\
$|\Psi_{12}^{+}\rangle$ & $Rev++$ & $Rev+-$ & $ID++$ & $ID+-$ \\
$|\Psi_{12}^{-}\rangle$ & $Rev+-$ & $Rev++$ & $ID+-$ & $ID++$ \\
\hline
\end{tabular}\label{out}
\end{center}
\end{table}

{\em Quantum Authentication}
\begin{itemize}
\item[(A.0)]  Authentication process begins when Alice asks Trent that
she wants to communicate with Bob.

\item[(A.1)]  Trent prepares an ordered set of $2N$ pairs of Bell state of
$|\Phi^{+}\rangle = \frac{1}{\sqrt {2}}(|00\rangle+|11\rangle)$. We
denote the $2N$ ordered EPR pairs as $[(P_{1}(T),
P_{1}(A)),(P_{2}(T), P_{2}(A)),$ $....,(P_{N}(T), P_{N}(A))]$ and
$[(P_{N+1}(T), P_{N+1}(B)),(P_{N+2}(T), P_{N+2}(B)),...., $
$(P_{2N}(T) , P_{2N}(B))]$.

Here the subscript indicates the ordering number of pairs, and $T$,
$A$, and $B$ represent the qubits of Trent, Alice and Bob,
respectively.

\item[(A.2)] Trent takes one qubit from each EPR pair,
say,$[P_{1}(T),P_{2}(T),...,P_{N}(T)]$
$([P_{N+1}(T),P_{N+2}(T),...,P_{2N}(T)])$ which is called the
$A(B)$-checking sequence, and keep it safely. The remaining sequence
of qubits $[P_{1}(A),P_{2}(A),...,P_{N}(A)]$
$([P_{N+1}(B),P_{N+2}(B),...,P_{2N}(B)])$ is called the
$A(B)$-authentication sequence.

\item[(A.3)]  Trent encodes $A(B)$-authentication sequence with
Alice's(Bpb's) identification numbers $ID(A)$ $(ID(B))$. If the
$i$-th value of $ID(A)$ is 1, Trent makes an Hadamard operation $H$
to $i$-th qubit of $A(B)$-authentication sequence. If it is 0,
identity operation $I$ is applied. The results of the operation on
$P_{i} (A)$ is $\{ (1-ID_{i}(A)) I + ID_{i}(A) H \} P_{i} (A)$.

\item[(A.4)]  Trent sends the $A(B)$-authentication sequences
$[P_{1}(A),P_{2}(A),...,P_{N}(A)]$,
$([P_{N+1}(B),P_{N+2}(B),...,P_{2N}(B)])$ to Alice(Bob).

\item[(A.5)] The legitimate user, Alice(Bob) knows her(his) $ID$ sequence.
She(he) decodes the $A(B)$-authentication sequence with her(his)
$ID$ sequence. The decoding method is the same as the Trent's
encoding method.  According to the $ID$ sequence, Alice(Bob) makes
an Hadamard operation $H$ or does nothing to the qubits of
$A(B$)-authentication sequence. By this decoding operation, the
qubits are restored to their original state.  Then she(he) measures
her(his) sequence in the $\sigma_{z}$ basis, and announces the
outcomes.

\item[(A.6)]  Trent measures the ordered $A(B)$-checking sequence and
compare the results with the Alice's(Bob's) results. If
Alice's(Bob's) result is the same as Trent's, then authentication
succeeded.  Otherwise, authentication failed and abort
communication.
\end{itemize}

{\em Quantum Direct Communication}
\begin{itemize}
\item[(C.1)]  Alice prepares a random sequence of
$M+n+q$ Bell states from two states,
$|\Phi^{+}\rangle_{T_{A}\,A}=\frac{1}{\sqrt{2}}(|00\rangle+|11\rangle)$
and
$|\Psi^{+}\rangle_{T_{A}\,A}=\frac{1}{\sqrt{2}}(|01\rangle+|10\rangle)$..
This random choice is Alice's secret information.  Bob prepares
$M+n+q$ Bell states of
$|\Phi^{+}\rangle_{T_{B}\,B}=\frac{1}{\sqrt{2}}(|00\rangle+|11\rangle)$..
The subscripts of the states represents who is going to keep them
after the process of (C.2).

\item[(C.2)]  Alice(Bob) takes one qubit from each pair and sends Trent the
ordered string of $M+n+q$ qubits which is named as the
$A$-sequence($B$-sequence) hereafter.  Alice(Bob) stores the
remaining ordered sequence of qubits in a safe place, which is named
as {\em the encoding sequence} ({\em the decoding sequence})
hereafter.

\item[(C.3)]  Alice randomly chooses $n$ checking positions of the ordered
encoding sequence and publicly announces it. Trent measures the $n$
checking qubits of the ordered $A$-sequence by using $\sigma_{z}$
basis and tells the outcomes to Alice. Alice measures the
corresponding qubits of the encoding sequence by using $\sigma_{z}$
basis and compares it with Trent's outcomes. She estimates error
rate and can detect a eavesdropper.  Bob's checking method is the
same as that of Alice and Trent.  They can detect eavesdropper on
the channel of Alice-Trent or Bob-Trent.

\item[(C.4)]  Trent performs Bell measurements on the qubits of the ordered
$A$ and $B$ sequences. In this Bell measurement, Trent does not have
to distinguish all of four different Bell states, but only needs to
distinguish $|\Phi^{\pm}\rangle$ and $|\Psi^{\pm}\rangle$ states.
After the Trent's measurement, the encoding sequence possessed by
Alice and the decoding sequence possessed by Bob became to be
entangled(Entanglement Swapping).  Trent sends his measurement
outcomes to Alice.

\item[(C.5)]  Alice receives the Trent's outcome and measures the encoding
sequence with $\sigma_{z}$ basis.  She randomly chooses $q$ checking
positions of the ordered encoding sequence and publicly announces
the positions. Bob performs measurement on the corresponding $q$
checking positions of the decoding sequence with $\sigma_{z}$ basis
and tells the outcome to Alice.  Alice can infer Bob's measurement
outcome from the effect of entanglement swapping, the information of
Trent's measurement outcome, her initial Bell state and her
measurement outcome.  Alice compares her inference with Bob's
corresponding announcements.  If there is no eavesdropper on the
line, their corresponding results should be correlated.  If there is
no correlation, the communication is aborted.  Table \ref{Cor} shows
the correlations.

\begin{table}
\caption{The correlation of entanglement swapping}
\begin{center}
\begin{tabular}{|c|c|c|c|c|} \hline
Bob's & Alice's & Trent's Bell & Alice's outcome & Bob's outcome
\\
initial state & initial state & measurement outcome & & \\ \hline
 & & $|\Phi^{\pm}\rangle$ & 0 & 0\\ \cline{4-5}      %2 line
 & $|\Phi^{+}\rangle$ & & 1 & 1 \\ \cline{3-5}     %3 line
 & & $|\Psi^{\pm}\rangle$ & 0 & 1 \\ \cline{4-5}     %4 line
$|\Phi^{+} \rangle$ & & & 1 & 0 \\ \cline{2-5}     %5 line
 & &  $|\Phi^{\pm}\rangle$ & 0 & 1 \\ \cline{4-5}     %6 line
 & $|\Psi^{+}\rangle$ & & 1 & 0 \\ \cline{3-5}     %7 line
 & &  $|\Psi^{\pm}\rangle$ & 0 & 0 \\ \cline{4-5}     %8 line
 & & & 1 & 1 \\ \hline
\end{tabular} \label{Cor}
\end{center}
\end{table}

\item[(C.6)] According to Alice's bit strings, she publicly announces the
positions that Bob needs to flip his measurement outcome on his
decoding sequence.  As she knows Bob's measurement outcome by using
entanglement swapping effect, she can send decoding information to
Bob.  Bob flips the value of his measurement outcome of the position
that Alice informed.  Then he can decode the her secret message.
\end{itemize}

For example, let's suppose that Alice prepares the ordered set of
Bell states $\{|\Phi^{+}\rangle, |\Phi^{+}\rangle,
|\Psi^{+}\rangle\}$, and Trent's Bell measurement outcomes are
$\{|\Phi^{+}\rangle, |\Psi^{+}\rangle, |\Psi^{+}\rangle\}$. When
Alice's measurement outcomes of the encoding sequence are \{0, 0,
1\}, Alice knows that Bob's outcomes must be \{0, 1, 1\}. Suppose
that Alice's secret message bit is $101$. According to the message,
Alice publicly announces $110$ which designates the positions Bob
needs to flip his measurement outcome. After the flipping, Alice's
message is transferred to Bob.

{\em Security analysis} - The proof of the security of our QDC
protocol is based on the security of the transmission of the $A$-
and $B$-sequence. The state of the transmitted qubits does not
contain any information of the secret message because they are
completely random and mixed.  The exposed information is just random
like that of coin flipping.  In our protocol, even Trent can not
know Alice's secret message since he doesn't know Alice's initial
state.

The qubit transmission and the checking method in our protocol is
similar to the procedure in BBM92 QKD protocol\cite{Bennett_3}.
Alice stores the encoding-sequence in her safe place, and Eve cannot
access it at all.  Therefore, the security of our protocol is the
same as that of the BBM92 QKD protocol.  The proof of security for
BBM92 protocol in ideal and practical conditions has been given
\cite{Ina,waks}. So our protocol is also unconditionally secure.

{\em Conclusion} - We have established the authenticated quantum
multiuser direct communication using entanglement swapping.  Its
security is the same as that of BBM92 protocol, which is
unconditionally secure.  The encoding of the message is processed
only after the authentication of the users and the confirmation of
the security of the quantum channel.  Our protocol, therefore, is
not in danger of exposure of information to Eve.  Furthermore the
leaked information to Eve is totally random, and does not contain
any information.

In this protocol we need only EPR paris.  It can be advantage in an
experiment.  The great feature of our protocol is that any two users
among $n$ subscribers can communicate each other. We don't need any
quantum channel linking two users, because the center, Trent,
connects two users Alice and Bob, and authenticates them.  This
structure is the same as that of nowadays telephone system, but its
security is much better than present technology.  It is
unconditionally secure.  Our scheme may be used for the safe
communication system.

\bibliographystyle{siam}

\end{document}